\documentclass[12pt]{amsart}

\usepackage{amsthm,amsmath,amssymb,amsrefs}

\usepackage{hyperref}

\usepackage{float}

\usepackage{color}

\usepackage{xcolor}

\usepackage{listings}

\usepackage{graphicx}

\counterwithin{figure}{section}

\newtheorem{problem}[figure]{Problem}

\usepackage[top=1in, bottom=1in, left=1in, right=1in]{geometry}

\newcommand{\Z}{\mathbb{Z}}

\newcommand{\F}{\mathbb{F}}

\author[]{Alexander Demin}
\address{Laboratoire d'informatique de l'{\'E}cole polytechnique, LIX, UMR 7161, CNRS, 1 rue Honor{\'e} d'Estienne d'Orves, 91120 Palaiseau, and Sorbonne Université, CNRS, 4 Pl. Jussieu, 75005 Paris, France}
\email{alexander.demin@lix.polytechnique.fr}

\author[]{Alexey Ovchinnikov}
\address{Department of Mathematics, Queens College, City University of New York, Queens, NY 11367 and
Ph.D. Programs in Mathematics and Computer Science, CUNY Graduate Center, 365 Fifth Avenue, New York, NY 10016}\email{aovchinnikov@qc.cuny.edu}\thanks{Research of the second and third authors was partially supported by a CUNY Institutional Research Grant Award}

\author[]{Vladimir Shpilrain}
\address{Department of Mathematics, The City College of New York, New York NY 10031,  and 
Ph.D. Programs in Mathematics and Computer Science, CUNY Graduate Center, 365 Fifth Avenue, New York, NY 10016
} \email{shpilrain@yahoo.com}

\begin{document}

%
%
%

\title{CHAMP: \lowercase{\uppercase{C}ayley \uppercase{HA}shing with \uppercase{M}atrix \uppercase{P}roducts}}

\begin{abstract}
Cayley hash functions hash binary strings by representing the bits of the message as elements of a semigroup and multiplying the corresponding elements. This gives Cayley hash functions useful homomorphic properties. Here, we propose CHAMP, a Cayley hash function based on products of $2\times 2$ matrices over a finite field, and discuss its security, implementation, and performance.
\end{abstract}

\maketitle

\section{Introduction}

Hash functions are easy-to-compute compression functions that take
a variable-length input and convert it to a fixed-length output.
Hash functions are used as compact representations, or digital
fingerprints, of data and to provide message integrity. {\it Cryptographic hash functions} have many information-security applications, notably in digital signatures, message authentication codes, and other forms of authentication. Cryptographic hash functions should satisfy the following basic security requirements:

\begin{enumerate}

\item {\it Collision resistance}: it should
be computationally infeasible to find two different inputs that hash
to the same output.
\medskip

\item {\it Preimage resistance} (sometimes called {\it non-invertibility}): it should be
computationally infeasible to find an input which hashes to a specified output.

\end{enumerate}

We note that collision resistance also implies {\it second preimage resistance}: it
should be computationally infeasible to find a second input that
hashes to the same output as a specified input.

A challenging problem is to determine mathematical properties of a
hash function that would ensure (or at least, make it likely) that
the requirements above are met.

 A direction that has been gaining momentum lately is using a pair of
elements, $A$ and $B$, of a semigroup $S$, to hash the ``0" and the ``1" bit, respectively. Then a bit string is hashed to a product of elements in the natural way.
For example, the bit string 1001011 will be hashed to  the element $BAABABB$.

Since hashing a bit string this way represents a walk on the Cayley
graph of the subsemigroup of $S$ generated by the elements $A$ and $B$, hash functions of this kind are often called {\it Cayley hash functions}. 
These hash functions have some useful functionalities that other hash functions (including the SHA family) do not have, see Section \ref{why}.

One might assume, in the spirit of ``there is no free lunch", that the price to pay for these functionalities should be weaker security. However, up to date all known attacks on Cayley hashing proposals have been {\it ad hoc}, i.e., have been using specific ``symmetry" in proposed pairs $(A, B)$ of semigroup elements.

In this paper, we propose a pair $(A, B)$ of matrices over $\F_p$ (for a safe prime $p$) with new properties that we argue should make the corresponding Cayley hash function immune to all known attacks.

The arrangement of the paper is as follows. We start by describing additional functionalities
of Cayley hash functions in Section \ref{why}. In  Section \ref{previous}, we describe some of the previous proposals for platform (semi)groups as well as our own new proposal. In Section \ref{formal}, we give a formal description of our hashing protocol and suggested parameters, as well as rationale for parameter selection. In Section \ref{Julia}, we give a Julia code for our protocol. In Section \ref{security}, we discuss security and possible attacks on our hash function. In Section \ref{performance}, we give performance data and compare performance of our hash function to that of SHA-512.  In Section \ref{NISTtests}, we report results of testing our hash function for pseudorandomness using the NIST statistical test suite. Finally, we conclude with some open problems in Section \ref{problem}.

\section{Additional functionalities of Cayley hash functions}\label{why}

Cayley hash functions have some very useful properties that standard hash functions, like SHA family, do not have. We explain these properties below. In what follows, $XY$ denotes  concatenation of bit strings $X$ and $Y$, and $H_{M}$ denotes the function that takes an arbitrary bit string to the corresponding product of matrices $A$ and $B$.
\begin{itemize}
\item Homomorphic property \begin{equation}\label{eq:hom}
H_{M}(XY)=H_{M}(X) \cdot H_{M}(Y)
\end{equation}
for any bit strings $X, Y$.
\item Associativity property \[H_{M}(XYZ)=H_{M}(XY) \cdot H_{M}(Z) = H_{M}(X)\cdot H_{M}(YZ)\] for any bit strings $X, Y, Z$.
\end{itemize}

Why are these properties useful? In contrast with hash functions in the SHA family, Cayley hash functions have the following useful functionalities:
\begin{enumerate}
\item To hash a bit string $X$, one does not have to read the whole $X$ first or even know the length of $X$; you can just hash $X$ ``as you go", bit by bit. You can even hash a stream of bits without knowing when it stops.
\item When an already hashed document $X$ is amended by $Y$, one does not have to hash the whole amended document $XY$ all over again, but rather hash just the amended part $Y$ and then multiply the result by the hash of the original document because~\eqref{eq:hom}.
Also note that one does not even have to know $X$ to compute $H_{M}(XY)$. Just knowing $H_{M}(X)$ is enough.

Moreover, one can replace any middle part of a hashed document as well. Suppose the original document is a concatenation of 3 bit strings: $XYZ$, and suppose one wants to replace $Y$ with $Y'$. Assuming $X$, $Y$, and $Z$ were already hashed before to $H_{M}(X)$, $H_{M}(Y)$, and $H_{M}(Z)$, respectively, we compute $H_{M}(Y')$, and then \[H_{M}(XY'Z)=H_{M}(X)H_{M}(Y')H_{M}(Z).\]
\item The property~\eqref{eq:hom} allows for parallel computations in a natural way.

It looks like all these features apply to the function $H_{M}$, the ``intermediate" function, and not to the actual hash function $H$, see Section \ref{actual}. Still, it is easy to see that all the features are preserved if the procedure of converting $H_{M}(X)$ to $H(X)$ is {\it deterministic, efficient, and efficiently invertible}. Here is how it works.

For example, if an already hashed document $X$ is amended by $Y$, here is how we can compute $H(XY)$ from $H(X)$ and $H(Y)$, respectively. First we recover $H_{M}(X)$ and $H_{M}(Y)$ from $H(X)$ and $H(Y)$.
This is possible since we assumed that converting $H_{M}(X)$ to $H(X)$ is efficiently invertible.
Then we compute~\eqref{eq:hom} and convert it to $H(XY)$.
\end{enumerate}

\section{Specific platform (semi)group}\label{previous}

While the high-level idea of Cayley hashing is definitely appealing, the choice of the platform semigroup $S$ and two elements $A, B \in S$ is crucial for security and efficiency. There have been many proposals based on matrix semigroups in $GL_2(\F)$ for various fields $\F$, in particular for $\F=\F_p$. This is because Cayley graphs of 2-generator semigroups in $GL_2(\F_p)$ often have a large girth as was shown by several authors, see e.g. \cite{BG,BSV,Masuda,H}. 
Cayley graphs of (semi)groups in $GL_n(\F_p)$ with $n >2$ have been considered, too (see e.g.~\cite{Battarbee}).

The first proposal of a Cayley hash function was due to Z\'emor \cite{Zemor}.
The matrices used, considered over $\F_p$, were \begin{displaymath}
A = \begin{pmatrix} 1 & 1 \\ 0 & 1 \end{pmatrix} , \hskip 0.5cm B = \begin{pmatrix} 1 & 0 \\ 1 & 1 \end{pmatrix}.
\end{displaymath}
This proposal was successfully attacked in \cite{TZ_attack} (finding collisions) and recently in   \cite{McKemmie} (finding preimages).
A specific property exploited in the attack in \cite{TZ_attack} was that this particular pair of matrices generates the whole semigroup of matrices from $SL_2(\Z)$ having nonnegative entries.
In addition, \cite{McKemmie} used the fact that this pair of matrices generates the whole {\it group} $SL_2(\Z)$.

The most cited proposal is what has become known as the Tillich-Z\'emor hash function \cite{TZ}. Their matrices were
\begin{displaymath}A = \begin{pmatrix} \alpha & 1 \\ 1 & 0 \end{pmatrix} , \hskip 1cm B = \begin{pmatrix} \alpha & \alpha+1 \\ 1 & 1 \end{pmatrix}.
\end{displaymath}
These matrices are considered over a field defined as
$R=\F_2[x]/(p(x))$, where  $\F_2[x]$ is the ring of polynomials over
$\F_2$,  $(p(x))$ is the ideal of $\F_2[x]$
generated by an irreducible polynomial $p(x)$ of degree $n$
(typically,  $n$ is a prime, $127 \le n \le 170$), and  $\alpha$ is a root of $p(x)$.

The reason for selecting such a ``fancy" field probably was to specifically avoid the attack in \cite{TZ_attack}. However, there were different attacks targeted at collision search in the Tillich-Z\'emor hash function, see  \cite{GIMS,  Petit_Rubik}. There were also attacks targeted at preimage search, see \cite{preimage}.

More recent proposals that use $2 \times 2$ matrices over $\F_p$ include \cite{BSV} and \cite{SosSH}. Matrices of larger size were used in \cite{AJS} and \cite{Battarbee}.

We, too, use a pair of $2 \times 2$ matrices, $A$ and $B$, over $\Z$  and then reduce the entries modulo a large prime $p$ to get matrices over $\F_p$, see Section \ref{parameters}. Rationale for selecting that particular pair is given in Section \ref{Rationale}.

If $p$ is a 256-bit prime, then a $2 \times 2$ matrix over $\F_p$ yields a 1024-bit string, the output of our hash function. If $p$ is a 128-bit prime, then we get a 512-bit string as the output.

\subsection{Converting output matrix to a bit string}\label{actual}

Denote by $H_{M}(X)$ the matrix over $\F_p$ obtained while hashing a bit string $X$. This is not a final output of hashing because the final output has to be a bit string of a fixed length. Denote the final output of hashing by $H(X)$.

The procedure for converting $H_{M}(X)$ to $H(X)$ has to be {\it deterministic, efficient, and efficiently invertible}. This is needed to preserve several useful features of the function $H_{M}(X)$, see Section \ref{why}.

\subsection{Converting a $2 \times 2$ matrix over $\F_p$ to a bit string of length 1024}\label{length1024}

If $p$ is a 256-bit prime, then each entry of $H_{M}(X)$ can be naturally converted to a bit string of length 256, and thus a $2 \times 2$ matrix over $\F_p$ can be naturally, by concatenating 4 bit strings of length 256, converted to a bit string of length 1024.

However, the way we suggest here is slightly different. We first invert each nonzero entry of $H_{M}(X)$ modulo $p$, and only after that convert to bit strings and concatenate.

This procedure is efficient; the slowest part is inverting modulo $p$. It is also deterministic and efficiently invertible.

\subsection{Converting a $2 \times 2$ matrix over $\F_p$ to a bit string of length 512}\label{length512}

If we want the output of a hash to be a 512-bit string, then we use a 128-bit prime $p$.

To convert $H_{M}(X)$ to a bit string of length 512, one does the same thing as in Section \ref{length1024}.

\section{ Algorithm description}\label{formal}

In the following description, we assume that $p$ is a 256-bit prime and the output of the hash function has 1024 bits. For a 512-bit output, we use a 128-bit prime, but the algorithm itself is the same.
\medskip

\noindent Given a bit string $X$ (the input):
\begin{enumerate}
\item Going left to right along the bit string $X$, compute the product of copies of $2 \times 2$ matrices $A$ and $B$ over $\F_p$, where the matrix $A$ corresponds to the ``0" bit and the
matrix $B$ corresponds to the ``1" bit. For example, if $X=10011$, compute the product $H_{M}(X)=BAABB$.
\item $H_{M}(X)$ is a $2 \times 2$ matrix whose entries are integers modulo $p$. Compute the inverse (in $\F_p$) of each nonzero entry of $H_{M}(X)$ and convert to a bit string of length 256. If an entry is 0, convert it to a string of 256 zeros.
Denote the entries of the obtained matrix by $h_{ij}$.
\item Concatenate the 4 bit strings $h_{ij}$ of length 256 obtained at Step 2 to a bit string of length 1024 in the following order: $h_{11}, h_{21}, h_{12}, h_{22}$.
The obtained bit string is $H(X)$, the output of our hash function.
\end{enumerate}

\subsection{Parameters}\label{parameters}

Recall that for a 1024-bit output of our hash function we use a 256-bit prime $p$, whereas for a 512-bit output we use a 128-bit prime.
The 256-bit prime number $p$ that we suggest is \[p = 2^{256} - 36113.\] This is a {\it safe prime}, i.e., $p-1 = 2q$ for another prime $q$. This is actually the largest 256-bit safe prime.
The 128-bit prime number $p$ that we suggest is \[p = 2^{128}  - 15449.\] This is the largest 128-bit safe prime.

A particular pair of matrices $(A, B)$ that we suggest is \[A= \begin{pmatrix} -2 & 1 \\ 4 & -3 \end{pmatrix},\quad   B = \begin{pmatrix} -5 & 2 \\ -6 & 2 \end{pmatrix}.\]
An alternative pair is \[A= \begin{pmatrix} -2 & 1 \\ 4 & -3 \end{pmatrix},\quad   B =\begin{pmatrix} 6 & 4 \\ -2 & -1 \end{pmatrix}.\]

\subsection{Rationale for parameter selection}\label{Rationale} 
\begin{enumerate}
\item The reason for selecting $p$ to be a safe prime is as follows. The determinant of both suggested matrices $A$ and $B$ is 2. Therefore, if there is a collision $u(A, B)=v(A, B)$ for two different products $u$ and $v$ of copies of $A$ and $B$, then $2^d = 1$ in the multiplicative group $\F_p^*$ of $\F_p$. Here $d$ is the difference in the lengths of
$u$ and $v$. 

Recall that the order of $\F_p^*$ is $(p-1)$.
If $p=2q+1$ is a safe prime, then $2^d = 1$ in $\F_p$ implies that $d$ is either 0 or at least $q$, i.e., very large. Thus, at least we have a guarantee that there are no collisions $u(A, B)=v(A, B)$ if $u(A, B)$ and $v(A, B)$ are not of the same length and are of length less than $\frac{p-1}{2}$. 
\item One of the reasons for selecting matrices $A$ and $B$ as in Section \ref{parameters} is for them to have determinant 2, thereby enforcing collision resistance, see above.

Another reason is that the suggested pairs of matrices should make the corresponding Cayley hash functions resistant to {\it length attack}, see the last two paragraphs of Section \ref{attacks}. For more details behind selecting these parameters, see Section \ref{security}, where security is discussed.
\end{enumerate}

\section{Implementations}\label{Julia}

We have implemented both CHAMP-1024 and CHAMP-512 in C code\footnote{Available at {\url{https://github.com/sumiya11/champ/tree/7cf76bcc5720a530151df5b8f90751cd3ecc6db9}}}. In all implementations, we used the pair of matrices \[A= \begin{pmatrix} -2 & 1 \\ 4 & -3 \end{pmatrix},\quad   B = \begin{pmatrix} -5 & 2 \\ -6 & 2 \end{pmatrix}.\]

Below is a 
basic code structure of CHAMP-1024 in Julia code that we provide for reference. Our fully functional optimized implementation of CHAMP-1024 (resp., CHAMP-512) has approximately 800 (resp., 600) lines of code.

\begin{lstlisting}[caption={A minimal reference implementation of \mbox{CHAMP-1024} in Julia.},label={fig:julia}]
const p = big(2)^256 - 36113    # Using "big" multi-precision integers
const A = big.([-2  1;  4 -3;])
const B = big.([-5  2; -6  2;])

function champ_1024(
        msg::Vector{UInt8},     # Input:  array of bytes, message
        nbits::Int,             # Input:  integer, message bit length
        digest::Vector{UInt8}   # Output: 128-byte buffer
    )

    S = big.([1  0; 0  1])                         # Identity matrix
    for i in 0:nbits-1
        bit = (msg[div(i,8)+1] >> (7 - i%8)) & 1   # The i-th bit of message
        M   = iszero(bit) ? A : B
        S   = mod.(S * M, p)                       # Mat.-mul. modulo a prime
    end

    inv0(x) = iszero(x) ? big(0) : invmod(x, p)    # Invert nonzero entries
    S = inv0.(S)

    vals = [S[1,1], S[2,1], S[1,2], S[2,2]]
    for k in 1:4                                   # Write 128 output bytes
        for j in 0:31
            digest[32*(k-1)+j+1] = UInt8((vals[k] >> (8*(31-j))) & 0xff)
        end
    end
end
\end{lstlisting}

\section{Security}\label{security}

Homomorphic properties of Cayley hash functions provide very useful additional functionalities (see Section \ref{why}), but intuitively, there should be a trade-off between additional functionalities and security (i.e., ``there is no free lunch").

For example, if for an input $Y$, the matrix $H_{M}(Y)$ is the identity matrix, then for any input $X$, one has
\[H_{M}(XY)=H_{M}(X)H_{M}(Y)=H_{M}(X),\] i.e., there is a collision. The only way to avoid this problem is to only have very long $Y$ such that $H_{M}(Y)=I$, the identity matrix. In particular, if
$A^k=I$ (or $B^k=I$), then $k$ has to be very large. This is the case with our proposed parameters. Indeed, the determinant of both $A$ and $B$ is 2, so if $A^k=I$, then $2^k=1$ in $\F_p$. Since $p$ is a safe prime, i.e., $p=2q+1$ for another prime $q$, $2^k=1$ implies here $k=q$, a 255-bit number. In other words, if $H_{M}(Y)=I$, then $Y$ is of length
$2^{255}$.

Still, the homomorphic property implies that if one collision was found, this would yield  multiple collisions. Indeed, if $H_{M}(Y)=H_{M}(Z)$, then $H_{M}(XY)=H_{M}(XZ)$ for any bit string $X$. On the other hand, over the 35-year history of Cayley hash functions, collisions have been found only in some ``highly symmetric" cases by using ad hoc attacks, see \cite{TZ_attack, GIMS,Petit_Rubik}.
In the absence of general methods for finding collisions for Cayley hash functions, the focus is shifting to security properties relevant to preimage resistance, see Section \ref{attacks} below.

\subsection{Security statements}\label{security  statement}

The following statements apply to classical as well as quantum computing settings.
\begin{enumerate}
\item There are provably no collisions $H(X)=H(Y)$ if the bit strings $X$ and $Y$ have different lengths and the difference between their lengths is less than $2^{255}$.

If $X$ and $Y$ are of the same length, we do not have a theoretical proof of the infeasibility of finding collisions $H(X)=H(Y)$, but experimental evidence suggests that lower bounds on the girth of the Cayley graphs of matrix (semi)groups found in the literature (\cite{BG, BSV, Masuda, H}) are ``too loose", i.e., the actual girth is actually huge in ``most" cases.
\item Our hash function has passed all NIST pseudorandomness tests. This is a good (albeit indirect) evidence of preimage resistance, but a stronger security property is infeasibility of recovering the first (or the last) bit of a bit string $X$ from the hash $H(X)$ with probability significantly larger than 0.5.

We claim that with our choice of parameters, it is infeasible to recover the first (or the last) bit of $X$ from $H(X)$ with probability significantly larger than 0.5 if $X$ has at least 1000 bits.
\end{enumerate}

\subsection{Possible attacks}\label{attacks}

For a while, the following property of semigroups of matrices over $\F_p$ has been considered an evidence of collision resistance. If two matrices $A$ and $B$, when considered over $\Z$, generate a {\it free semigroup}, then there cannot be any collisions of the form $u(A, B)=v(A, B)$ modulo $p$ unless at least some of the entries in $u(A, B)$ or $v(A, B)$ are greater than $p$.

Thus, several authors were looking for pairs of matrices $A$ and $B$ over $\Z$ whose products $u(A, B)$ would exhibit the slowest growth of maximal entries. Equivalently, the Cayley graph of the semigroup generated by $A$ and $B$ over $\F_p$ should have a large {\it girth}, see \cite{BG,H}.
The pair \[A= \begin{pmatrix} 1 & 1 \\ 0 & 1 \end{pmatrix},   \quad B = \begin{pmatrix} 1 & 0 \\ 1 & 1 \end{pmatrix}\] was considered in the seminal paper \cite{Zemor} where Cayley hash functions were introduced. The problem with this pair, however, is that these two matrices generate the whole semigroup of invertible $2 \times 2$ matrices over $\Z$ with nonnegative entries. This often makes it feasible to find a preimage of a given matrix product by using a procedure similar to the Gaussian elimination, only without divisions (see \cite{TZ_attack} for details).  

In \cite{BSV}, the pair \[A= \begin{pmatrix} 1 & 2 \\ 0 & 1 \end{pmatrix},\quad   B = \begin{pmatrix} 1 & 0 \\ 2 & 1 \end{pmatrix}\]
was suggested. It was shown that, if $p$ is a 256-bit prime, there cannot be any collisions of the form $u(A, B)=v(A, B)$ modulo $p$ if the length of both $u(A, B)$ and $v(A, B)$ is less than 203.

However, this may not seem to be such a big selling point of the corresponding hash functions.
First of all, it is understood that, in general (i.e., for any hash function), hashing very short bit strings is not preimage resistant since a preimage in that case can be found just by brute force, i.e. by hashing all possible bit strings up to a given length and comparing the results. In \cite{Mullan}, the authors offered an efficient algorithm for finding collisions of length $O(\log p)$ for some Cayley hashing proposals based on matrices from $SL_2(\F_p)$.
This is still rather short given that $p$ is typically on the order of $2^{128}$  for $2^{256}$.
However, because of the homomorphic property~\eqref{eq:hom}, finding a short collision entails finding long collisions as well, which is why provably avoiding short collisions (as in \cite{BSV}) is still important for Cayley hash functions. More about this is in Section~\ref{problem}.

Second and more interesting, having a pair $(A, B)$ ``too symmetric", as in the proposals of
\cite{Zemor} and \cite{BSV}, might make the corresponding hash function vulnerable to a {\it length attack}. The high-level idea is that multiplying (on the right) by the matrix $A$ increases elements in the second column, whereas multiplying (on the right) by the matrix $B$ increases elements in the first column. Thus, by looking at the product matrix, one can ``locally" guess which matrix was the rightmost factor and therefore which bit was the rightmost in the input bit string. This attack was shown to have a success rate of over 50\% for input bit strings of length up to a couple of thousand bits. Of course, this is nowhere near recovering a whole preimage, but since this is essentially the only known ``generic" attack against Cayley hash functions based on matrices, it has to be addressed.

Informally, the reason why this attack may work is that the magnitude of entries in a matrix product increases monotonically. Then, given a matrix product, one can multiply it by $A^{-1}$ or $B^{-1}$ to see which one would decrease the magnitude of entries. Of course, there is no ``magnitude" in $\F_p$, but this might work ``locally" if $p$ is large and the input bit string is relatively short.

There are two ways to foil such an attack. The first way is rather obvious: if both matrices $A$ and $B$ have determinant 2, then $A^{-1}$ and $B^{-1}$ have determinant $2^{-1}$, which is a very large number in $\F_p$, so multiplying a matrix by either $A^{-1}$ or $B^{-1}$ will most likely increase the magnitude, not decrease.

Another way to foil the length attack is more sophisticated; it is based on an observation from \cite{Lyapunov}. The observation is: ``on average" (in a precise sense), entries in a random product of matrices $A$ and $B$ (over $\Z$) grow as a linear combination of $\lambda_1^n$  and $\lambda_2^n$, where $\lambda_1$ and $\lambda_2$ are eigenvalues of the ``average matrix" $M=\frac{1}{2}(A+B)$. If at least one of  $\lambda_i$ is negative, then the growth (as a function of $n$) of their linear combination may not be monotone, and indeed it is not, as computer experiments show.

Now let us look at one of the pairs of matrices that we suggest: \[A= \begin{pmatrix} -2 & 1 \\ 4 & -3 \end{pmatrix},\quad   B = \begin{pmatrix} -5 & 2 \\ -6 & 2 \end{pmatrix}.\] For this pair, \[M=\frac{1}{2}(A+B) = \begin{pmatrix} -3.5 & 1.5 \\ -1 & -0.5 \end{pmatrix},\] and the eigenvalues of this matrix are $-2 \pm \frac{\sqrt{3}}{2}$, so they are both negative.

The alternative pair of matrices that we suggest: \[A= \begin{pmatrix} -2 & 1 \\ 4 & -3 \end{pmatrix},\quad   B = \begin{pmatrix} 6 & 4 \\ -2 & -1 \end{pmatrix}.\] Here \[M=\frac{1}{2}(A+B) = \begin{pmatrix} 2 & 2.5 \\ 1 & -2 \end{pmatrix},\] and the eigenvalues of this matrix are $\pm \sqrt{\frac{13}{2}}$, so one of them is positive and the other one is negative.

Figures 1 and 2 show how the ``length" of a matrix product changes if one uses the matrices \[A= \begin{pmatrix} -2 & 1 \\ 4 & -3 \end{pmatrix},\quad  B = \begin{pmatrix} -5 & 2 \\ -6 & 2 \end{pmatrix}.\] By the ``length" here we mean the sum of the absolute values of the matrix entries. We see that the change is clearly not monotonic in this case.

\begin{figure}\label{figure1}
    \centering
    \includegraphics[width=0.8\linewidth]{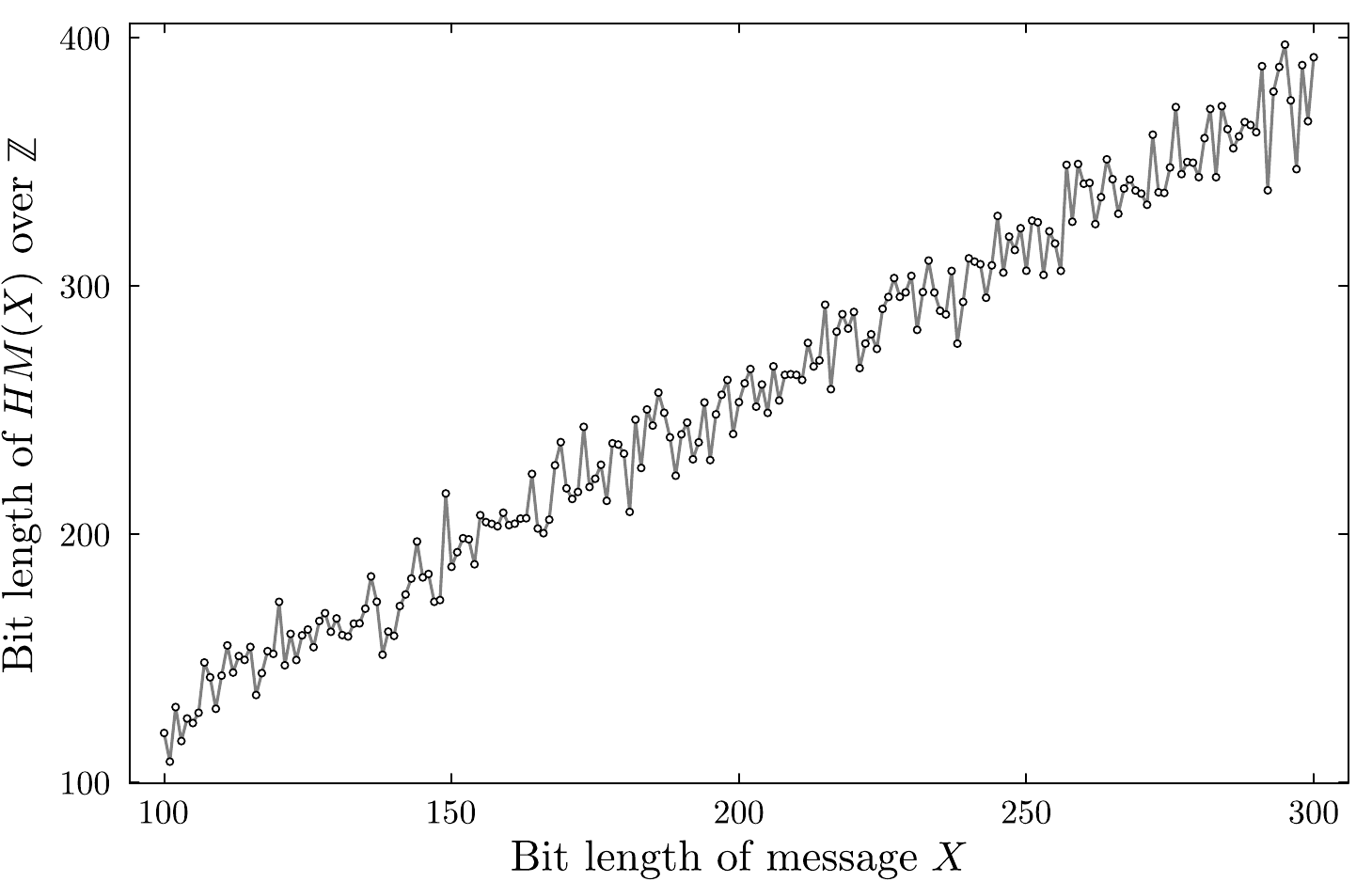}
    \caption{Bit length of the sum of absolute values of the entries of $H_{M}(X)$, the latter computed over $\mathbb{Z}$ without modular reduction, as a function of the bit length of the input $X$. The inputs $X$ are chosen uniformly at random.}

\vskip 2cm
%
%
    \includegraphics[width=0.8\linewidth]{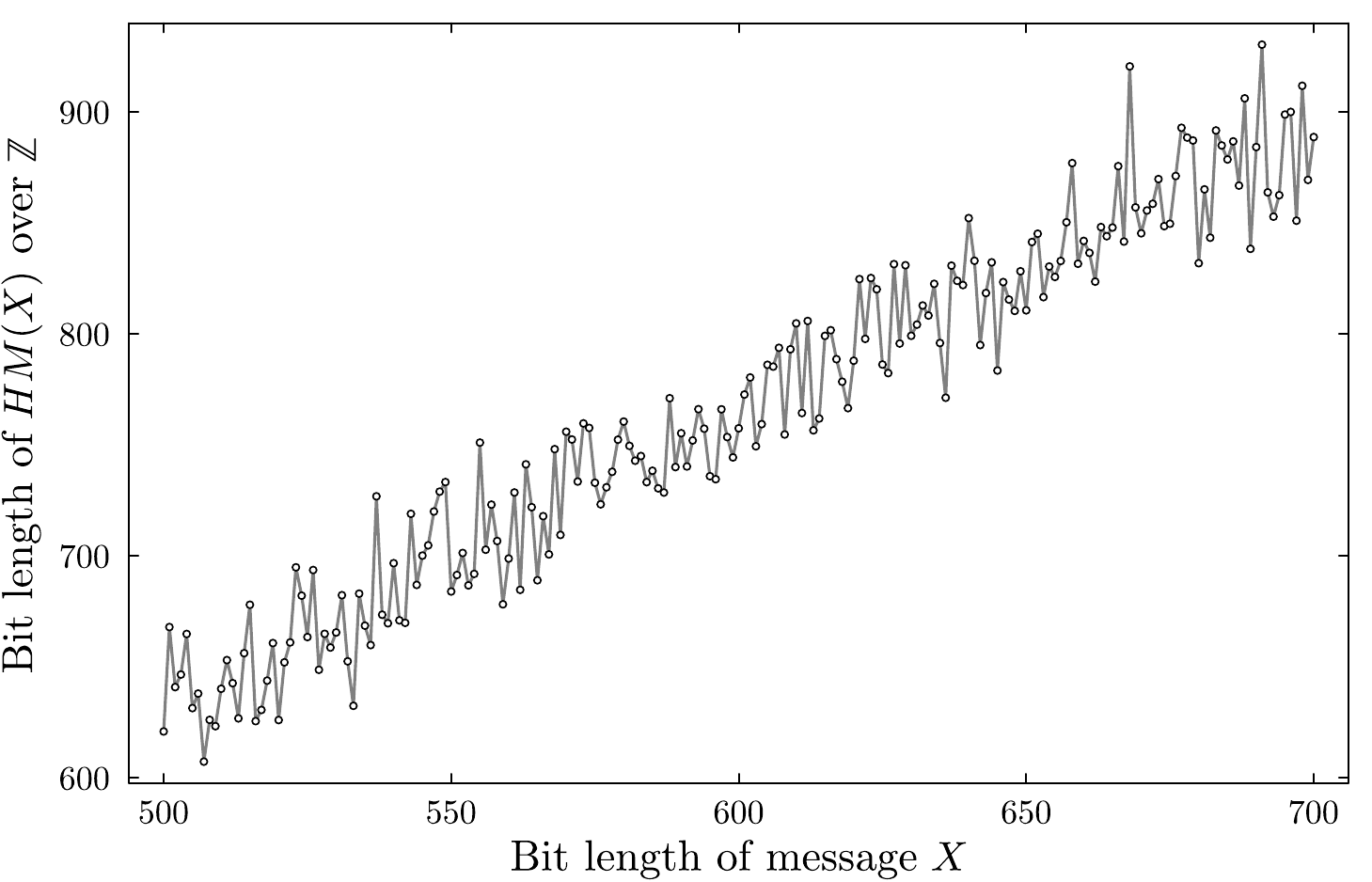}
    \caption{Bit length of the sum of absolute values of the entries of $H_{M}(X)$.}
    \label{fig:placeholder}
\end{figure}

\subsection{Birthday attack}\label{attacks2}

In \cite{Yuan}, the authors notice that, since both our matrices $A$ and $B$ have determinant 2, any product of length $n$ will have determinant $2^{n}$. Therefore, all fixed-length outputs lie in one ``determinant fiber" of size $p(p^2-1)$ rather than the full matrix space of size $p^4$. 

Then they offer a theoretical analysis of a Birthday attack in this case and show that, upon accumulating \[\sqrt{p(p^2-1)} \approx 2^{192}\] hash evaluations of CHAMP-512 (respectively, $\approx 2^{384}$ hash evaluations of CHAMP-1024), one can find collisions with probability at least 0.39.

This, of course, does not make collision search feasible under the current level of mankind's computational abilities, but given that there are no known serious attacks, theoretical or experimental, even in a ``simple" case where \[A= \begin{pmatrix} 1 & 2 \\ 0 & 1 \end{pmatrix},\quad B = \begin{pmatrix} 1 & 0 \\ 2 & 1 \end{pmatrix}\] (see Section \ref{problem}), the Birthday attack is still something to keep in mind.

\section{Performance}\label{performance}

We implemented our hash function in the 512-bit and 1024-bit output versions, each in two variants: reference and optimized implementations.

The optimized implementation incorporates two main performance boosters. First, since the entries of $A$ and $B$ are small integers, we accumulate the entries of several consecutive matrix products using native machine integer types before reducing modulo $p$, thus using fewer multiple-precision arithmetic operations. Second, we use a look-up table that stores the hash value corresponding to each possible byte ($256$ entries in total); equivalently, the table contains all possible products of eight generators (each being either $A$ or $B$). This allows the implementation to process the input byte-by-byte rather than bit-by-bit.

In this section, we report experimental performance of the optimized implementation.

\begin{table}[h]
\centering
\small
\begin{tabular}{lccc}
\hline
Setup & CHAMP-512 & CHAMP-1024 & SHA-512 \\
\hline
\hyperref[perf:setup-1]{(1)} i5-8250U Windows & 28.12 (MB/s) & 17.77 (MB/s) & 282.43 (MB/s) \\
\hyperref[perf:setup-2]{(2)} i9-13900 Linux & 49.78 (MB/s) & 44.03 (MB/s) & 919.05 (MB/s) \\
\hline
\end{tabular}
\caption{Throughput comparison across machine setups. Entries report the median MB/s for 100 MB input chunks.}
\label{table:perf}
\end{table}

Our benchmarking was performed in two different setups representing popular platforms:
\begin{enumerate}
    \item \label{perf:setup-1} An x86-64 laptop running Windows 10 on an Intel i5-8250U CPU, with 1.80 GHz processor frequency reported by the system for the benchmark session. The CHAMP-512 and CHAMP-1024 were compiled from source using GCC 11.1.0 with the flags \texttt{-O3 -DNDEBUG}. The SHA-512 baseline uses the compiled \texttt{bcrypt} library accessible via the Win32 CNG interface. The CHAMP-512 and CHAMP-1024 implementations run in single-threaded mode.
    \item \label{perf:setup-2} An x86-64 workstation running
	Linux 6.1.0-1029-oem on an Intel i9-13900 CPU. No fixed processor frequency was recorded for the benchmark session; \texttt{/proc/cpuinfo} snapshots taken during the benchmark showed per-core instantaneous readings up to about 4.60 GHz on active cores. The implementations were compiled
	with GCC 11.4.0 using the same release flags
	\texttt{-O3 -DNDEBUG}. The CHAMP benchmarks again use our
	optimized single-threaded C implementations. The \mbox{SHA-512} baseline
	uses the OpenSSL software interface through \texttt{libcrypto}.
\end{enumerate}

Performance was measured by repeatedly hashing a 100 MB chunk of data and collecting the running times. Table~\ref{table:perf} reports the resulting throughput in MB/s.

We observe that CHAMP-512 is faster than CHAMP-1024, in part because it uses 128-bit integers instead of the 256-bit ones for implementing arithmetic operations in $\F_p$.

As for comparing CHAMP-512 to SHA-512, Table 1 shows that CHAMP-512 appears to be 10 to 20 times slower. However, due to the homomorphic property (see Section \ref{why}), the input can be split in, say, 8 parts to be hashed simultaneously, and then the resulting 8 matrices would be multiplied together before converting to a 512-bit string. This would speed up computation almost by the factor of 8.

\section{NIST pseudorandomness tests}\label{NISTtests}

We have tested the outputs of our hash function for pseudorandomness using the NIST statistical test suite~\cite{NIST} for both $512$-bit and $1024$-bit output cases.


We describe the experimental setup for the case of $1024$-bit output; the $512$-bit one is similar.
For a positive number $X$ smaller than $2^{1024}$, we denote by $H(X)$ the result of applying our hash function to the $1024$-bit little-endian binary representation of $X$.
We compute $H(1),H(2),\ldots,H(N)$ and concatenate the results in a single bit string of length $1024 N$ bits. We pick $N$ sufficiently large so that the length of the resulting bit string is at least $100$ MB.

We run the tests with the default values of parameters, using $100$ bit streams each $10^6$ bits in length.
The default values of parameters were:
\begin{verbatim}
    [1] Block Frequency Test - block length(M):         128
    [2] NonOverlapping Template Test - block length(m): 9
    [3] Overlapping Template Test - block length(m):    9
    [4] Approximate Entropy Test - block length(m):     10
    [5] Serial Test - block length(m):                  16
    [6] Linear Complexity Test - block length(M):       500
\end{verbatim}

Testing results for both pairs of matrices suggested in Section \ref{parameters}  are collected in four text files\footnote{Available at {\scriptsize\url{https://github.com/sumiya11/champ/tree/7cf76bcc5720a530151df5b8f90751cd3ecc6db9/NIST}}}. For each pair of matrices, tests were run for both 512-bit and 1024-bit output, which is why we have four files with test results altogether.

According to these results, our hash function $H(X)$ has successfully passed all NIST pseudorandomness tests, for both pairs of matrices suggested in Section \ref{parameters}.

\section{Open problems}\label{problem}

As we mentioned in Section \ref{attacks}, it was shown in \cite{BSV} that the girth of the Cayley graph of the semigroup generated by $A= \left(\begin{array}{cc} 1 & 2 \\ 0 & 1 \end{array} \right)$ and $B = \left(\begin{array}{cc} 1 & 0 \\ 2 & 1 \end{array} \right)$ (considered over $\F_p$) is at least 203 if $p$ is a 256-bit prime.

The fact that the corresponding Cayley hash function has not been successfully attacked up to date probably means that the actual girth of the Cayley graph here is a lot larger than 203. Therefore we ask:

\begin{problem}
What is the actual girth of the Cayley graph of the semigroup generated by $A= \left(\begin{array}{cc} 1 & 2 \\ 0 & 1 \end{array} \right)$ and $B = \left(\begin{array}{cc} 1 & 0 \\ 2 & 1 \end{array} \right)$ (considered over $\F_p$) if $p = 2^{256} - 36113$?
\end{problem}

The prime $p = 2^{256} - 36113$ is a {\it safe prime}, i.e., $p-1 = 2q$ for another prime $q$. This is actually the largest 256-bit safe prime.


We note that if negative entries are allowed, the situation becomes a lot more mysterious.
We ask:

\begin{problem}
What is the actual girth of the Cayley graph of the semigroup generated by $A= \left(\begin{array}{cc} 1 & 2 \\ 0 & 1 \end{array} \right)$ and $B = \left(\begin{array}{cc} 1 & 0 \\ -2 & 1 \end{array} \right)$ (considered over $\F_p$) if $p = 2^{256} - 36113$?
\end{problem}

With the help from Nicola Guglielmi \cite{Guglielmi} we were able to get a lower bound on the
girth of the Cayley graph in this case, following the method used in \cite{BSV}. The lower bound is 270 and again, we believe that the actual girth is a lot larger.

\medskip

We note that since $p$ in these problems is a 256-bit number, a $2 \times 2$ matrix over $\F_p$ is representable by a bit string of 1024 bits. Therefore, by the pigeonhole principle, an upper bound on the girth of the Cayley graph in this case is 1025+1025=2050.
\medskip

As for pairs of matrices we suggest in Section \ref{parameters}, we could, of course, ask the same question, but there are currently no methods to even try to attack such a problem for those
matrices.

\end{document}